# Density Fluctuations and Energy Spectra of 3D Bacterial Suspensions

Zhengyang Liu[1], Wei Zeng[1,2], Xiaolei Ma[1], and Xiang Cheng[1,*]

[1] Department of Chemical Engineering and Materials Science, University of Minnesota, Minneapolis, MN 55455, USA

[2] College of Life Science and Technology, Guangxi University, Nanning 530004, Guangxi, China

[*] Corresponding author: Email: xcheng@umn.edu

Giant number fluctuations are often considered as a hallmark of the emergent nonequilibrium dynamics of active fluids. However, these anomalous density fluctuations have only been reported experimentally in two-dimensional dry active systems heretofore. Here, we investigate density fluctuations of bulk *E. coli* suspensions, a paradigm of three-dimensional (3D) wet active fluids. Our experiments demonstrate the existence and quantify the scaling relation of giant number fluctuations in 3D bacterial suspensions. Surprisingly, the anomalous scaling persists at small scales in low-concentration suspensions well before the transition to active turbulence, reflecting the long-range nature of hydrodynamic interactions of 3D wet active fluids. To illustrate the origin of the density fluctuations, we measure the energy spectra of suspension flows and explore the density-energy coupling in both the steady and transient states of active turbulence. A scale-invariant density-independent correlation between density fluctuations and energy spectra is uncovered across a wide range of length scales. In addition, our experiments show that the energy spectra of bacterial turbulence exhibit the scaling of 3D active nematic fluids, challenging the common view of dense bacterial suspensions as active polar fluids.

# Introduction

Density fluctuations are intrinsic to equilibrium systems at finite temperatures and control diverse physical phenomena ranging from the liquid-gas phase transition, to radiation scattering, and to the compressibility of materials.[1] For isotropic homogeneous fluids in equilibrium, the number of particles in a volume $V$ fluctuates around the average number $N$ with a standard deviation $\Delta N$ proportional to the square root of the mean number $\sqrt{N}$ (or equivalently the square root of the volume $\sqrt{V}$). However, this well-established relation dictated by the central limit theorem may not hold when the system falls out of equilibrium. Composed of large collections of self-propelled particles, active fluids are a class of nonequilibrium soft materials, which display many fascinating behaviors beyond the expectation of equilibrium thermodynamics.[2-9] In particular, an active fluid can exhibit anomalously large density fluctuations, the so-called giant number fluctuations (GNF), where $\Delta N$ grows substantially faster than $\sqrt{N}$.[10-21] Such anomalous density fluctuations may arise simply from the formation of dense clusters in motility-induced phase separation,[22] but could also be the profound consequence of activity-driven hydrodynamic instabilities that couple emergent collective flows with density fluctuations.[4] The mechanism of the latter has triggered extensive research interests in recent years, motivating our current experimental investigation.

GNF have been observed in a broad range of living and non-living active fluids such as vibrated granular rods,[10-13] swarming bacteria[14-16] and mammalian cells,[17,18] self-propelled cytoskeleton,[19] and synthetic colloidal swimmers.[20,21] As a result, GNF are often viewed as a hallmark of the emergent nonequilibrium dynamics of active fluids. However, although significant progress has been made in the theoretical understanding of GNF over the past two decades,[23-33] systematic experiments that can quantitatively verify theoretical and numerical predictions are still few and far between. Particularly, previous experiments on GNF all focused on dry two-dimensional (2D) or quasi-2D active systems,[10-21] where steric interparticle and particle-boundary interactions dominate system dynamics.[4] These system-specific interactions lead to a variety of the scaling relations of density fluctuations quantified generically by $\Delta N/\sqrt{N} \sim N^\alpha$. Scaling exponents $\alpha$ ranging from 0.13 up to 0.5 have been reported in different experiments. Note that $\alpha = 0$ for isotropic homogeneous equilibrium systems obeying the central limit theorem, whereas the upper bound $\alpha = 0.5$ corresponds to a system with maximum density fluctuations. In contrast, quantitative measurements of GNF in 3D wet active fluids, where hydrodynamics govern the interparticle interactions and conserve the total momentum of systems, have not been achieved heretofore. Important questions including how the long-range hydrodynamic interaction and the dimensionality of systems affect the density fluctuations of active fluids have yet to be addressed experimentally. Such experiments would be of value for revealing the intricate coupling between emergent active flow and density fluctuations and elucidating the origin of giant number fluctuations in 3D wet active fluids.

Here, we present our systematic experiments on the density fluctuations of bulk bacterial suspensions, a premier example of 3D wet active fluids.[4,8] We first examine the spatiotemporal density correlations and measure the correlation length and time of density fluctuations with increasing bacterial concentrations. Moreover, we demonstrate the existence of GNF and quantify the scaling of density fluctuations in bulk bacterial suspensions. Surprisingly, we find that the scaling of GNF persists at small scales in low-concentration suspensions well before the transition to active turbulence with collective bacterial swimming. Such an unusual behavior likely stems from the long-range hydrodynamic interaction of 3D wet active fluids. To explore the origin of 3D density fluctuations, we further inspect the energy spectra of suspension flows, which characterize the scale-dependent flow structures of active turbulence. Interestingly, the scaling of the energy spectra extracted from our experiments shows the signature of active nematics and challenges the common view of dense bacterial suspensions as active polar fluids. Finally, our methodical measurements reveal a scale-invariant density-independent coupling between GNF and energy spectra of bacterial suspensions across a wide range of length scales. We show that the same coupling also governs the transient kinetics of a bacterial suspension during its transition towards active turbulence. Taken together, our study provides benchmark experiments on density fluctuations and energy spectra of bulk bacterial suspensions and sheds light on the origin of giant number fluctuations in 3D wet active fluids.

## Experiment

We use a genetically engineered light-powered *Escherichia coli* strain as our model active particles (Appendix A). In a typical experiment, a bacterial suspension of control volume fraction $\phi$ is injected into a sealed chamber of 20 mm by 3 mm by 140 μm. Here, $\phi = nV_b$, where $n$ is the number density of bacteria and $V_b$ is the volume of a single bacterium. We estimate $V_b = \pi(w_b/2)^2 l_b \approx 1$ μm$^3$, where $l_b = 3$ μm and $w_b = 0.65$ μm are the average length and width of bacterial body. Without external oxygen supply, bacteria quickly consume all the oxygen in the chamber and stop swimming after about 5 minutes. We then illuminate the suspension using a high-intensity light, which saturates the light response of bacteria even in the dense region of high-$\phi$ suspensions used in our experiments. Thus, bacteria show a density-independent average swimming speed of $v_0 = 15 \pm 3$ μm/s. A video of the suspension is taken 50 μm above the bottom wall of the chamber by a bright-field inverted microscope at a frame rate of 30 fps with a field of view of 420 μm by 360 μm (Fig. 1A) and an estimated depth of field of 6 μm. We use a standard Particle Image Velocimetry (PIV) algorithm to extract the 2D in-plane velocity field $\boldsymbol{v} = (v_x, v_y)$ in the 3D suspension (Appendix A). Since bacteria are used as tracers, PIV tracks bacterial flow, instead of fluid flow. Nevertheless, it has been shown that fluid flow and bacterial flow share quantitatively similar statistical structures.[34] Our PIV analyses reveal the intermittent vortices and jets of collective bacterial swimming in high-$\phi$ suspensions (Fig. 1B and Supplementary Video 1), a phenomenon often referred to as active turbulence.[35,36]

Note that the smallest dimension of our system is $H = 140$ μm, which is about two orders of magnitude larger than the body size of single bacteria. Recent numerical and experimental studies have shown that the dynamics of bacterial suspensions exhibit bulk behaviors at such a large scale.[37-39] Away from system boundaries, the flow of bacterial suspensions is isotropic. Hence, imaging a 2D plane within a bulk suspension yields representative results and provides all the necessary statistical information on the density fluctuations and flow structures of bulk 3D suspensions.

It is challenging to count the number of bacteria in a 3D dense suspension of fast moving bacteria directly. Fortunately, by virtue of the Beer-Lambert law, local bacterial density is monotonically correlated with the local intensity of microscope images, where darker regions correspond to higher bacterial densities (Figs. 1A, C and Supplementary Video 1). Such a correlation has been exploited in previous experiments in probing the dynamics of bacterial suspensions and actin filaments.[19,40,41] To calibrate the density-intensity correlation, we prepare bacterial suspensions of different $\phi$ and image the suspensions under the same illumination (Fig. 1D inset). The mean image intensity decreases with increasing $\phi$ following an approximately linear relation (Fig. 1D), agreeing with the Beer-Lambert law for samples of small thickness and weak absorptivity appropriate to our experiments. The linear density-intensity relation is independent of the selection of the field of view down to the size of single pixels.

## Results

**Spatiotemporal correlations of density and velocity fluctuations.** The monotonic correlation between image intensity and bacterial density allows us to map the spatiotemporal patterns and measure the correlation length and time of the density fluctuations of 3D bacterial suspensions. Specifically, we compute the density spatial correlation, $C_n$, and the density temporal autocorrelation, $C_{t,n}$, for suspensions of different $\phi$ (Figs. 2A and B) and compare them with the corresponding velocity spatiotemporal correlations, $C_v$ and $C_{t,v}$, extracted from PIV (Figs. 2C and D). The definition of $C_n$, $C_{t,n}$, $C_v$ and $C_{t,v}$ are given in Supplementary Materials (SM) Sec. 1.1. The correlation length $\lambda$ and time $\tau$ are determined when the corresponding normalized correlation function decays to $1/e$.

Figure 2E shows the density correlation length $\lambda$ at different $\phi$, which quantifies the scale of density inhomogeneities in suspensions. $\lambda$ is small at low $\phi$, gradually increases with $\phi$ and reaches a plateau of ~

$5l_b$ above $\phi_c$ in high-$\phi$ suspensions with active turbulence, where $\phi_c$ is the transition concentration to active turbulence. $\phi_c = 4.05\%$ in our experiments, quantitatively agreeing with our previous experiments as well as the prediction of kinetic theories.[42] The velocity correlation length follows a qualitatively similar trend and also saturates above $\phi_c$, consistent with previous experiments.[43] The saturated density correlation length is about one-half of the saturated velocity correlation length. Our previous study on high-$\phi$ bacterial suspensions in a planar Couette cell with an adjustable gap thickness has shown that the velocity correlation length is linearly proportional to the size of the system,[38] which is set by the minimum dimension of the sealed chamber $H = 140$ μm in our current study.

The correlation times of velocity and density fluctuations, $\tau$, show more interesting trends with increasing $\phi$ (Fig. 2F). While the velocity correlation time increases with $\phi$ when $\phi < \phi_c$, the density correlation time decreases with increasing $\phi$. The two time scales are well separately for low-$\phi$ suspensions. These opposite trends can be understood as follows. The local velocities of low-$\phi$ suspensions are small and change directions rapidly due to the random tumbling of bacteria within PIV boxes. As a result, the velocities decorrelate quickly. With increasing $\phi$ towards active turbulence, local spatiotemporal correlations between bacteria build up,[37,44,45] which lead to longer persistence of the direction of local velocities and therefore longer velocity correlation times. The effect culminates at $\phi_c$, resulting in the longest velocity correlation time, reminiscent of the critical slowing down in continuous phase transitions of equilibrium systems.[1] Above $\phi_c$, the lifetime of the intermittent vortices and jets of active turbulent flow is approximately constant independent of $\phi$,[43] giving a constant velocity correlation time ~ $6\tau_b$. Here, $\tau_b = l_b/v_0 = 0.2$ s is the characteristic swimming time scale of *E. coli*. Different from the velocity temporal autocorrelation, the density temporal autocorrelation depends on the magnitude, instead of the direction, of local velocities. Large velocities give rise to a fast temporal variation of local densities. Since the magnitude of velocities increases monotonically with $\phi$, the density correlation time continuously decreases, approaching ~ $4\tau_b$ at high $\phi$. As both the density and velocity fluctuations are controlled by the active turbulence above $\phi_c$, the two time scales become comparable in high-$\phi$ suspensions. A simple scaling argument can be formulated based on the above picture, which provides a good estimate of the density correlation time in the low-$\phi$ and high-$\phi$ limit (SM Sec. 1.2).

**Giant number fluctuations.** Exploiting further the linear intensity-density relation (Fig. 1D), we examine the possible existence of GNF in 3D bacterial suspensions by calculating the standard deviation of bacterial number $\Delta N$ for subsystems of different sizes (SM Sec. 2). Figure 3A shows $\Delta N$ as a function of the area of subsystems $l^2$ for bacterial suspensions of different $\phi$, where $l$ is the side length of square subsystems. Since the depth of field of our images $d$ is fixed, the average number of bacteria in the subsystems of volume $l^2 d$ is $N = \phi l^2 d / V_b$. Thus, $N$ is linearly proportional to $l^2$ at a fixed average $\phi$ and $\sqrt{N} \sim l$. To highlight the deviation from the central limit theorem, we normalize $\Delta N$ by $l$ (or equivalently $\sqrt{N}$). A system obeys the central limit theorem without GNF should show a constant $\Delta N/l$, independent of subsystem size $l$.

At any given $\phi$, $\Delta N/l$ of bacterial suspensions indeed plateaus following the central limit theorem when $l$ is a few times larger than the density correlation length at the corresponding $\phi$, $\lambda(\phi)$ (Fig. 3A). The central limit theorem applies at large length scales owing to the spatial average over multiple dense and dilute regions. Nevertheless, when $l$ is comparable or smaller than $\lambda(\phi)$, bacterial suspensions of all $\phi$ in our study exhibit obvious GNF, where $\Delta N/l$ increases with increasing $l^2$. The result demonstrates the existence of GNF in 3D bacterial suspensions. Remarkably, $\Delta N/l$ versus $l^2$ converges to the same scaling in the small $l$ limit even for low-$\phi$ suspensions without active turbulence. Such a finding is in contrast to 2D active systems, where GNF diminishes continuously with decreasing particle concentrations and disappears completely for low-concentration samples.[10-12,14,19] Although GNF at small scales has been reported in one 2D system, where gliding myxobacteria form moving colonies on agar substrates,[15] the origin of the density correlation at small scales in this 2D system is not clear. For 3D bacterial suspensions, kinetic theories have predicted that bacteria show spatiotemporal correlations at low concentrations well before the transition to active turbulence due to the long-range hydrodynamic interaction.[37,44] Although these theories focused on the effect of the correlations on the enhanced diffusion of passive tracers and velocity fluctuations in dilute suspensions, our experiments show that such correlations also lead to GNF at small length scales in low-$\phi$ bacterial suspensions.

The strength of GNF can be quantified by the scaling exponent $\alpha$ in the scaling relation $\Delta N/\sqrt{N} \sim N^\alpha$. We extract $\alpha$ by fitting the experimental data with power-law functions at $l < \lambda(\phi)$. The inset of Fig. 3A shows $\alpha$ as a function of $\phi$, where $\alpha$ remains approximately constant at $0.30 \pm 0.03$ for all $\phi$. For high $\phi \geq 5.6\%$, $\alpha$ stabilizes to $0.33 \pm 0.01$. The result quantitatively agrees with the theoretical prediction of $\alpha = 1/3$ for 3D suspensions of polar-ordered self-propelled particles with a long-range hydrodynamic interaction.[25] However, the GNF scaling shown in Ref. [25] is for ordered suspensions without viscous damping. The theory predicts that the ordered phase of bacterial suspensions at low Reynolds numbers ($Re$) is convectively unstable. Kinetic theories developed for low-$Re$ 3D active suspensions do show the emergence of collective bacterial swimming with large-scale bacterial orientational order,[46-49] where density fluctuations arise in the nonlinear regime beyond the linear stability analysis.[47] Nevertheless, quantitative investigation of GNF in low-$Re$ 3D wet active fluids has not been conducted heretofore. Our experiments not only measure the GNF scaling in bulk bacterial suspensions, but also show that the scaling persists in low-concentration suspensions at small scales well before the emergence of the long-range bacterial orientational order. Note that although the accuracy of the power-law fitting is limited due to the finite length scale of our experiments especially for low-$\phi$ suspensions, the existence of the scale-dependent GNF in bacterial suspensions of different concentrations is robust and does not rely on the fitting.

Lastly, we also analyze the scale-dependent density fluctuations in terms of the Fourier transform of the Ursell function, $S(\mathbf{k}) = \mathcal{F}(S_{nn}(\mathbf{r}))$ (Appendix A), where the Ursell function $S_{nn}(\mathbf{r})$ quantifies the spatial correlation of density fluctuations $S_{nn}(\mathbf{r}) = \langle \delta n(\mathbf{r} + \mathbf{r_0}, t)\delta n(\mathbf{r_0}, t)\rangle_{\mathbf{r_0}, t}$.[1] Here, $\delta n(\mathbf{r}, t) = n(\mathbf{r}, t) - \langle n \rangle$ is the density fluctuation at $\mathbf{r}$ and $t$ around the mean density $\langle n \rangle$, whereas the standard deviation of local density fluctuations is $\Delta n(\mathbf{r}) = \langle \delta n(\mathbf{r})^2 \rangle_t$. For isotropic liquids, $S(\mathbf{k})$ is equivalent to the static structure factor when $\mathbf{k} \neq 0$, which is intimately linked to $\Delta N$ in a subsystem of size $l$ considered above via[18]

$$\frac{\Delta N}{\sqrt{N}} = \frac{1}{l}[\int d^2\mathbf{k} \, |\hat{\chi}_A(\mathbf{k})|^2 S(\mathbf{k})]^{1/2}. \tag{1}$$

Here, $|\hat{\chi}_A(\mathbf{k})|^2 = l^4 \text{sinc}^2\left(\frac{k_x l}{2}\right)\text{sinc}^2\left(\frac{k_y l}{2}\right)$ is the Fourier spectrum of the characteristic function of the square region of area $l^2$. As bacterial suspensions are isotropic, we consider only the orientational average $S(k) = \langle S(\mathbf{k}) \rangle_{|\mathbf{k}|=k}$ (Fig. 3B). Consistent with the density fluctuations in the real space, $S(k)$ plateaus at small $k$, indicating the normal density fluctuation at large scales dictated by the central limit theory. The decrease of $S(k)$ at the intermediate and large $k$ suggests the existence of GNF at the intermediate and small length scales. In particular, $S(k)$ of different $\phi$ converge at large $k$, showing the abnormal density fluctuations at small scales, independent of the average concentration of bacterial suspensions.

**Energy spectra.** To understand the origin of scale-dependent GNF in bulk bacterial suspensions, we further investigate the flow structures of bacterial suspensions. Similar to GNF, the velocity field of bacterial suspensions also exhibits scale-dependent structures, which are usually characterized by its energy spectrum, $E(k)$ (Appendix A).[50-54] $E(k)$ measures the kinetic energy density at different scales in terms of wavenumber $k = 2\pi/l$, which is related to the mean kinetic energy density via $\langle v^2 \rangle/2 = \langle v_x^2 + v_y^2 \rangle/2 = \int_0^\infty E(k)dk$. Figure 4 shows $E(k)$ of bacterial suspensions at different $\phi$. In the dilute suspension of $\phi = 0.8\%$, $E(k)$ is independent of $k$ in the small $k$ limit and decreases at high $k$. The oscillation observed at high $k$ likely arises from PIV errors due to the small number of bacteria in each PIV boxes at low $\phi$. With increasing $\phi$, the plateau of $E(k)$ at small $k$ increases sharply. In the turbulent regime above $\phi_c$, the kinetic energy concentrates at scales much larger than the size of single bacteria, even though the flow is driven by the swimming of individual bacteria. The overall trend of $E(k)$ with increasing $\phi$ qualitatively agrees with large-scale particle simulations.[31,55]

The energy spectra of low-$\phi$ suspensions with uncorrelated pusher swimmers have been predicted,[55]

$$E(k) = 4\pi n \kappa^2 \left[\frac{1}{3} + \frac{\cos(kl_d)}{(kl_d)^2} - \frac{\sin(kl_d)}{(kl_d)^3}\right] \frac{\varepsilon^4 k^2}{l_d^2} K_2^2(k\varepsilon), \quad (2)$$

where $\kappa$ is the dipole strength and $l_d$ is the dipolar length of *E. coli*. $\varepsilon$ is the distance for the regularization of the dipolar flow field. $K_2$ is the modified Bessel function of the second kind. The fitting of Eq. (2) agrees reasonably well with our experiments at low $\phi$ in the small $k$ limit (Fig. 4). Particularly, Eq. (2) predicts that $E(k)$ is independent of $k$ as $k \to 0$, a key feature confirmed by our experiments. A simple dimensional analysis shows that the plateau $E(k)$ at the small $k$ follows $\lim_{k \to 0} E(k) \sim n\kappa^2$ for uncorrelated swimmers of number density $n$. Here, the dipole strength can be estimated as $\kappa = Fl_d/\eta = \xi v_0 l_d/\eta = 300.8$ μm³/s, where $\eta$ is the viscosity of the buffer. $\xi$ is the drag coefficient of a bacterial body along its major axis, which can be calculated based on the geometry of bacterial body $\xi = 3\pi\eta w_b[1 - (1 - l_b/w_b)/5]$.[56] $l_d = 1.9$ μm is taken from direct measurements.[57] Thus, $\lim_{k \to 0} E(k) \approx 7 \times 10^2$ μm³/s, within the same order of magnitude of our experiments. The discrepancy between Eq. (2) and experiments at large $k$ arises from PIV errors and the local bacterial correlations as shown by density fluctuations.

We also extract the scaling exponent $\beta$ of the energy spectra, $E(k) \sim k^{-\beta}$, by fitting $E(k)$ at intermediate $k$, where a significant change of $E(k)$ with $\phi$ occurs and $E(k)$ exhibits good power-law relations. $\beta$ increases with $\phi$ and saturates around 3 at high $\phi > \phi_c$ (Fig. 4 inset). The saturated scaling exponent quantitatively agrees with previous experiments on the active turbulence of high-concentration sperm suspensions and *B. subtilis* suspensions at large $k$.[51,52] However, $E(k)$ in Ref. [51] increases with $k$ at small $k$ and shows a non-monotonic trend with a pronounced peak. Such a discrepancy is likely attributed to the confined geometry used in [51], which limits the size of turbulent vortices[38] and causes the increase of $E(k)$ with $k$ when $k$ is smaller than the reciprocal of the vortex scale. The large system size of our experiments allows us to probe energy spectra at small $k$ without the influence of system boundaries. Remarkably, our experiments show $E(k) \sim k^0$ in the small $k$ limit and $\sim k^{-3}$ at large $k$, exhibiting the universal scaling predicted for 3D wet active *nematic* fluids.[36,58,59] This finding, in combination with our previous results on the linear relation between the velocity correlation length and the system size,[38] challenges the popular view of dense bacterial suspensions as active *polar* fluids.[4,34,36,51] Theories of wet active nematic fluids predict a long-wavelength hydrodynamic instability with the most unstable mode at $k \to 0$.[4,25,46-49] As a result, the velocity correlation length is set by the size of the system. In contrast, the phenomenological theories of active polar fluids predict an intrinsic vortex scale independent of system sizes,[34,51,60] which dictates a constant velocity correlation length at which energy spectra peak.

**Density fluctuation–flow coupling in steady state.** Both GNF and energy spectra probe the scale-dependent dynamic structures of bacterial suspensions. While the former measures density fluctuations at different scales, the latter probes flow energies contained at different scales. Although both quantities have been extensively studied, a direct correlation between the two has not been explicitly considered heretofore. Indeed, the trends of GNF and energy spectra show similar characteristics at high $\phi$, both exhibiting a rapid increase at small scales and plateaus at large scales (Figs. 3 and 4). Such a similarity suggests that a density-independent correlation between density fluctuations and kinetic energy may exist in the $k$ space. To verify the hypothesis, we plot the kinetic energy density $E(k)$ against the density fluctuations $\Delta N$ at each corresponding scale $l = 2\pi/k$ (Fig. 5A). All the $\Delta N$-$E(k)$ pairs in the turbulent regime fall onto a master curve over a wide range of length scales extending from the size of a single PIV step (~ 8 μm) up to the size of the entire system (> 140 μm), regardless of the volume fractions of the samples. In contrast, $\Delta N$-$E(k)$ scatters strongly for low-$\phi$ suspensions without active turbulence. It is certainly not surprising that density fluctuations positively correlate with kinetic energy in general, as both are increasing functions of bacterial density. However, the collapse of data of high-$\phi$ suspensions of different volume fractions at different scales is quite unexpected. The result demonstrates a scale-invariant and density-independent coupling between density fluctuations and kinetic energy density: $E(k)$ of the same value gives rise to quantitatively similar $\Delta N$ regardless length scales and volume fractions. The relation further suggests that the couplings between density fluctuations and flow energy at different scales are independent. A similar trend can be also observed when plotting $E(k)$ versus $S(k)$ (Fig. 5B), although the collapse is not as good as that shown in $\Delta N$-$E(k)$ for suspensions close to the transition concentration $\phi_c \approx 4.0\%$.

Density and velocity fluctuations are connected via the continuity equation at the mean field level,[47]

$$\frac{\partial n}{\partial t} + \boldsymbol{u} \cdot \nabla n = -\nabla \cdot (n\bar{v}_0 \boldsymbol{d}), \tag{3}$$

where $n(\boldsymbol{x},t)$ and $\boldsymbol{u}(\boldsymbol{x},t)$ are the bacterial density field and the fluid velocity field, respectively. $\boldsymbol{d}(\boldsymbol{x},t)$ is a unit vector field, indicating the local average orientation of bacteria. $\bar{v}_0(\boldsymbol{x},t)$ is the local average bacterial swimming speed, which may depend on the local density $n$. Diffusion is typically small compared to advection and thus neglected. Although looking simple, Eq. (3) is highly nonlinear with strong coupling between $n$, $\bar{v}_0$, $\boldsymbol{u}$ and $\boldsymbol{d}$ via kinetic equations. While the linear stability analysis of the coupled equations has been implemented on both the isotropic and fully-aligned states of suspensions, density remains uniform in the linear regime.[47,49] Hence, the rise of density fluctuations is an intrinsically nonlinear phenomenon, requiring the full solution of the kinetic equations in the turbulent state. As a result, a direct quantitative relation between velocity fields and density fluctuations of active turbulence cannot be easily obtained from Eq. (3) without recourse to experimental or numerical insights. The challenge is similar to that faced by the infamous problem of high-$Re$ turbulence,[61] even though active turbulence is driven by the swimming of bacteria with $Re \sim 10^{-4}$. Saintillan and Shelley showed numerically that the local density $n$ anti-correlates with the divergence of the relative bacterial flux $\nabla \cdot (n\bar{v}_0 \boldsymbol{d})$ for 2D active turbulence.[47] Here, we experimentally investigate the correlation between density fluctuations and flow structures of 3D active turbulence. Different from the results of the 2D numerical simulation, we find that the correlation between local density and bacterial flux is too weak to be detected (crosses in Fig. 6B). Instead, we show that density fluctuations and active turbulent flow are coupled in the $k$ space, where a scale-invariant density-independent correlation between density fluctuations and energy spectra is uncovered over a wide range of length scales.

The coupling between density fluctuations and energy spectra at each length scale can be qualitatively understood as follows. By integrating over a box of a side length $l$ and applying the divergence theorem on Eq. (3), we achieve

$$\frac{\partial N_l}{\partial t} = \oiint [n(\bar{v}_0 \boldsymbol{d} + \boldsymbol{u})] \cdot d\boldsymbol{S} = \oiint n\boldsymbol{v} \cdot d\boldsymbol{S}, \tag{4}$$

where $N_l$ is the total number of bacteria in the box and $\partial N_l/\partial t$ represents the density fluctuations in the box. $d\boldsymbol{S}$ is the differential surface area of the box. $\boldsymbol{v} = \bar{v}_0 \boldsymbol{d} + \boldsymbol{u}$ is the velocity of bacterial flow measured by PIV. The incompressibility of the fluid flow $\nabla \cdot \boldsymbol{u} = 0$ is applied when deriving Eq. (4). For simplicity, we consider the initial growth of density fluctuations starting from the turbulent state established in the linear regime with spatially uniform density. $n$ is then a constant that can be taken outside the integration. If the local turbulent flow has a correlation length much larger than $l$, then the velocity at the one side of the box would be strongly correlated with the velocity at the opposite side of the box. Bacteria are simply advected through the box without changing the number of bacteria inside. Thus, the flow at the large scales does not contribute to the density fluctuations inside the box. Similarly, if the correlation length is much smaller than $l$, the spatial average of randomly orientated small-scale turbulent vortices over the surface area $l^2$ would lead to zero density flux at the surface and, therefore, does not contribute to the density fluctuations inside the box either. Thus, the density fluctuations inside the box of length $l$ should be dominated by the flow with a correlation length on the order of $l$. Because $E(k = 2\pi/l)$ is the Fourier transform of the two-point velocity-velocity spatial correlation and quantifies the strength of bacterial flow at the length scale $l$,[61] $E(k = 2\pi/l)$ naturally correlates with the density fluctuations at the scale $l$ as shown in Fig. 5. The simple argument above assumes a one-way coupling, where the turbulent flow triggers density fluctuations and not vice versa. This simple assumption is valid only at early times during the initial growth of density fluctuations from the state of uniform density in the linear regime. As density fluctuations build up, the back influence of density inhomogeneities on the turbulent flow may not be ignored. Thus, our experimental finding provides not only a quantitative understanding of the origin of density fluctuations in dense bacterial suspensions, but also important insights into the nonlinear dynamics of 3D active turbulence.

To illustrate the correlation between density fluctuations and kinetic energy at a given scale in real space, we calculate the correlation of *local* density fluctuations and kinetic energy at the smallest length scale of our PIV analysis, i.e., the step size of PIV at $l = 2.75 l_b$ (see SM Sec. 3.1 for the definition). The local density fluctuation $\Delta n(r, t)$ and kinetic energy $E(r, t)$ are extracted from the image intensity field and the PIV velocity field, respectively (Fig. 6A). The normalized correlation between $\Delta n(r, t)$ and $E(r, t)$ averaged over all $r$ and $t$ is then computed at different $\phi$ (disks in Fig. 6B). At low $\phi$ without active turbulence, the correlation is weak and fluctuates around zero. It then increases sharply with $\phi$ as bacterial suspensions transition to active turbulence. A constant positive correlation persists in the turbulent regime when $\phi > \phi_c$, which is substantially larger than the correlation between local density and the divergence of bacterial flux (see SM Sec. 3.2 for the definition). The results provide a concrete example of the coupling between density fluctuations and kinetic energy at small scales.

As a comparison, we also calculate the direct correlation between the local density (instead of local density *fluctuations*) and kinetic energy (squares in Fig. 6B)(see SM Sec. 3.3 for the definition). Local density and kinetic energy show small but positive correlations when $\phi < \phi_c$. A high bacterial density enhances local bacterial alignment and therefore increases local flow velocity, a feature similar to that observed in 2D bacterial swarming on agar substrates.[14] In contrast, local density and kinetic energy become anti-correlated in active turbulence above $\phi_c$. The finding is reminiscent of the preferential accumulation of passive inertial particles in high-*Re* turbulence,[62] where heavy particles concentrate at convergence zones at which multiple vortices interact to form a saddle-like flow of low flow velocity. Nevertheless, since the particle Stokes number of bacteria is $St \sim 10^{-5}$, the inertial effect that drives particle preferential accumulation in high-*Re* turbulence cannot be present in bacterial suspensions. Instead, the anti-correlation may arise from the unique features of bacterial locomotion in shear flow such as rheotaxis and shear-induced trapping of bacteria.[63,64] More works are certainly needed to reveal the origin of the density-flow correlation in active turbulence, which are beyond the scope of our current study focusing on density fluctuations and will be the subject of future investigations.

**Density fluctuation–flow coupling in transient state.** The same scale-invariant density-independent correlation between density fluctuations and energy spectra is also present in the kinetic process during the bacterial turbulent transition. Taking the advantage of the light-powered bacteria, we trigger the onset of bacterial turbulence by suddenly turning on the light illumination on high-$\phi$ bacterial suspensions at $t = 0$.[42]

Figure 7A shows the temporal evolution of density fluctuations and turbulent flow in a bacterial suspension of $\phi$ = 6.4% through this kinetic process (see also Supplementary Video 2). At $t$ = 0, the local flow velocities are close to 0. The suspension shows no sign of density fluctuations. At $t$ = 40 s, although the magnitudes of velocities are still small, local alignment of velocity directions can be clearly observed, displaying the characteristic pattern of turbulent vortices and jets. Weak density fluctuations start to emerge. At $t$ = 103 s, the magnitude of velocities grows substantially and saturates. The suspension reaches the steady-state active turbulence with strong density fluctuations. Note that the response time of individual bacteria to light triggering is much shorter than the emergence of collective flows. A single bacterium recovers its swimming speed within a few seconds after turning on the light.[42]

Quantitatively, we monitor the temporal evolution of density fluctuations and energy spectra of the suspension in the transient state before the suspension reaches the steady-state turbulence (SM Sec. 2.4). Figure 7C shows the growth of GNF during the turbulent transition. Different from the steady-state GNF at different $\phi$, where strong GNF persist at small length scales even for low-$\phi$ suspensions, the high-$\phi$ bacterial suspension shows no or very weak GNF at all scales at the onset of active turbulence near $t$ = 0 with a small scaling exponent $\alpha \approx 0.10$. The strength of GNF, quantified by $\alpha$, gradually increases over time (the black line in Fig. 7B). $\alpha$ saturates to the steady-state value of 0.33 above $t \approx 90$ s. Interestingly, the growth of GNF is significantly delayed compared with the formation of collective turbulent flow, which is measured by the area fraction of the regions with strong local velocity alignment, $M$.[42,65] $M$ reaches a plateau at a much earlier time ~ 30 s (the blue dotted line in Fig. 7B). The finding provides experimental evidence on a key prediction of the kinetic theories of active suspensions,[46-49] where density fluctuations are shown to be the consequence of the nonlinear development of the hydrodynamic instability of pusher swimmers and appear only at long times. In the linear regime at the onset of the hydrodynamic instability, although the collective turbulent flow has already established, density remains uniform. While $\alpha$ and $M$ show a clear separation of time scales, the growths of $\alpha$ and the kinetic energy $E$ are strongly correlated and exhibit quantitatively similar trends in the transient state (the black and the orange lines in Fig. 7B), demonstrating the coupling between density fluctuations and flow kinetic energy in the kinetic process.

Lastly, we also analyze the correlation between $\Delta N$ and $E(k)$ during the turbulent transition of bacterial suspensions of different $\phi$ (Fig. 7D). When $\phi > \phi_c$, all our data at different $\phi$ and $k$, as well as at different $t$ except for the earliest time of 10 s close to the light response time of bacteria ("×" markers in Fig. 7D), collapse into the same master curve obtained from the steady-state measurements. The result shows that the $k$-space kinetic energy density controls not only the steady-state GNF but also the rise of GNF at the corresponding scale.

## Discussion and conclusions

We have conducted systematic experiments investigating density fluctuations and energy spectra of 3D bulk bacterial suspensions over a wide range of concentrations in both the steady and transient states. We examined the density and velocity spatiotemporal correlations and uncovered a dynamic slowing down as the system approaches the bacterial turbulent transition. More importantly, we demonstrated the existence of giant number fluctuations in bulk bacterial suspensions and quantified their scaling relation at different concentrations. We found that the same scaling persisted at small scales even in low-concentration suspensions well before the transition to active turbulence. Moreover, we also measured the energy spectra of bacterial suspensions of different concentrations and showed the spectral properties of the active turbulence of dense bacterial suspensions in the bulk limit. The scaling of the energy spectra of our bulk samples suggests that dense bacterial suspensions behave like an active nematic fluid, rather than an active polar fluid. Lastly, by comparing density fluctuations and energy spectra at different scales, we revealed a universal correlation between density fluctuations and kinetic energies across two orders of magnitude of length scales stretching from the scale of single bacteria up to the size of the system. We further showed that such a scale-invariant and density-independent correlation also dominated the kinetic process during the transition towards active turbulence and governed the rise of density fluctuations in 3D wet active fluids.

In addition to these previously unknown features, our study also verified several important theoretical predictions on 3D wet active fluids:

*i*. Our study provided experimental evidence on the existence of local spatiotemporal correlations between bacteria in 3D dilute suspensions well before the transition to active turbulence.[37,44]

*ii*. The measured energy spectra of low-concentration bacterial suspensions confirmed the key feature of the energy spectra of uncorrelated pusher swimmers in the small wavenumber limit,[55] whereas the scaling of the energy spectra of high-concentration suspensions at both small and large wavenumbers testified the universal scaling predicted for 3D wet active nematics.[36]

*iii*. The delayed onset of density fluctuations uncovered in our experiments supported the central prediction of the kinetic theories on the nonlinear development of the hydrodynamic instability of pusher suspensions.[46-49]

It is worth of noting that there have been extensive studies on GNF and energy spectra of active fluids as two independent topics. Our experiments showed that GNF and energy spectra are coupled in 3D wet active fluids following a scale-invariant density-independent correlation. The finding can potentially help to coordinate independent research efforts in understanding these two different emergent properties of active turbulence.

## Appendix A: Materials and methods

**Light-powered *E. coli*.** We introduced a light-driven transmembrane proton pump, proteorhodopsin (PR), to wild-type *E. coli* (BW25113) by transforming the bacteria with plasmid pZE-PR encoding the SAR86 γ-proteobacterial PR-variant.[66] The activity of PR was correlated with the intensity of light. Thus, we could control the swimming speed of bacteria using light of different intensities. In our experiments, we used high-intensity light, which saturated the light response of bacteria. The average swimming speed of bacteria was fixed at $v_0 = 15 \pm 3$ μm/s.

The bacteria were cultured at 37 °C with a shaking speed at 250 rpm for 14-16 hours in terrific broth (TB) [tryptone 1.2% (w/w), yeast extract 2.4% (w/w), and glycerol 0.4% (w/w)] supplemented with 0.1 g/L ampicillin. The culture was then diluted 1:100 (v:v) in fresh TB and grown at 30 °C for 6.5 hours. PR expression was triggered by supplementing the culture medium with 1 mM isopropyl *β*-D-thiogalactoside and 10 μM ethanolic all-trans-retinal in the mid-log phase, 3 hours after the dilution. The bacteria were harvested by gentle centrifugation (800*g* for 5 min). After discarding the culture medium in the supernatant, we resuspended bacteria with DI water. The resuspended suspension was then centrifuged again at 800*g* for 5 min, and finally adjusted to the target concentration for experiments.

**Sample preparation and microscopy.** To prepare the sample for microscopy, we constructed a seal chamber made of glass slides (25 mm by 75 mm) and #1.5 coverslips (18 mm by 18 mm). We first glued (NOA 81, Norland, NJ) two coverslips on a glass slide, side-by-side, leaving a 3-mm separation between the two coverslips. We then covered the 3-mm separation with another coverslip to form a channel. We pipetted bacterial suspensions into the channel. Finally, the two ends of the channel were sealed by UV glue (NOA 76, Norland, NJ) to form a sealed chamber.

Images of the bacterial suspensions were taken 50 μm above the bottom surface of the sealed chamber by a Nikon Ti-E inverted microscope in the bright-field mode using a 20× (NA 0.5) objective. The field of view was 420 μm by 360 μm. All videos were recorded at 30 frames per second using an Andor Zyla sCMOS camera.

**Particle image velocimetry (PIV).** Two-dimensional in-plane velocity fields were extracted by Particle Image Velocimetry (PIV) analysis using the openPIV package in Python (Liberzon et al., OpenPIV). We fixed the box size to be 16.5 μm, which was larger than the size of a single bacterial body but smaller than the velocity correlation length. A step size of the half of the box size with $\Delta x = 2.75 l_b = 8.25$ μm was used by convention, which set the spatial resolution of the velocity fields.

**Fourier transform of the Ursell function.** The Fourier transform of the Ursell function was calculated as follows. First, for a given frame of time, we measured the local light intensity $I(r)$ at the spatial resolution of PIV by averaging pixel intensities over non-overlapping square boxes of size $\Delta x = 8.25$ μm. The fluctuations of the local light intensity were then calculated as $\delta I(r) = I(r) - \langle I \rangle$, where $\langle I \rangle$ is the average light intensity over all the boxes at different $r$. Since the local light intensity is linearly proportional to the local bacterial density $n$, we have $\delta I(r) \sim \delta n(r)$. We then used the built-in Fast Fourier Transform (FFT) function of Python numpy.fft package to convert the discrete density fluctuation field $\delta n(r) = \delta n(x,y)$ to the density fluctuation field in the momentum space $\delta n(k) = \delta n(k_x,k_y)$. Note that to compensate the difference between FFT and the continuous Fourier transform, we multiplied the $k$-space density fluctuation from FFT by the square of the spatial spacing of the discrete $n(r)$, $\Delta x^2 = 68.1$ μm². The Fourier transform of the Ursell function in the $k$-space was then computed as

$$S(k_x, k_y) = \frac{1}{2A} \langle \delta n(k_x, k_y) \delta n^*(k_x, k_y) \rangle,$$

where $A$ is the total area of the field of view and $^*$ denotes the complex conjugate. $\langle \cdot \rangle$ represents an average over multiple images from different times. Finally, the orientational average $S(k)$ was obtained by averaging all $S(k_x,k_y)$ with $k = (k_x^2 + k_y^2)^{1/2}$ falling in the bin $[k, k+dk]$.

**Energy spectra.** A similar method was also used to calculate the energy spectra of bacterial suspensions. First, we applied the Python package numpy.fft to convert the discrete velocity field $v(r) = [v_x(x,y), v_y(x,y)]$ obtained from PIV to the velocity field in the momentum space $v_k(k) = [u_k(k_x,k_y), v_k(k_x,k_y)]$. To compensate the difference between FFT and the continuous Fourier transform, we multiplied the $k$-space velocity from FFT by the square of the spatial spacing of the discrete $v(r)$, $\Delta x^2 = 68.1$ μm². The point-wise kinetic energy density in the $k$-space was then computed as

$$E(k_x, k_y) = \frac{1}{2A} \langle u_k(k_x, k_y) u_k^*(k_x, k_y) + v_k(k_x, k_y) v_k^*(k_x, k_y) \rangle,$$

where $A$ is the total area of the field of view and $^*$ denotes the complex conjugate. $\langle \cdot \rangle$ represents an average over multiple images from different times. There was a slight difference in computing the orientational average of $S(k)$ and $E(k)$. Specifically, the energy spectrum $E(k)$ was obtained by first averaging all $E(k_x,k_y)$ with $k = (k_x^2 + k_y^2)^{1/2}$ falling in the bin $[k, k+dk]$ and then timing the average with $2\pi k$. The procedure ensures that $\langle \boldsymbol{v}^2 \rangle/2 = \langle v_x^2 + v_y^2 \rangle/2 = \int_0^\infty E(k)dk$. Thus, the method is mathematically equivalent to the calculation of $E(k)$ from the Fourier transform of the two-point velocity correlation function $\langle \boldsymbol{v}(\boldsymbol{r_0}) \cdot \boldsymbol{v}(\boldsymbol{r_0} + \boldsymbol{r}) \rangle$.[61]

## Acknowledgements

We thank D. Ghosh, Y. Peng, Y. Qiao and K. Zhang for the help with experiments and fruitful discussions. The research is supported by NSF CBET 1702352, NSF CBET 2028652 and the David and Lucile Packard Foundation.

**Figures:**

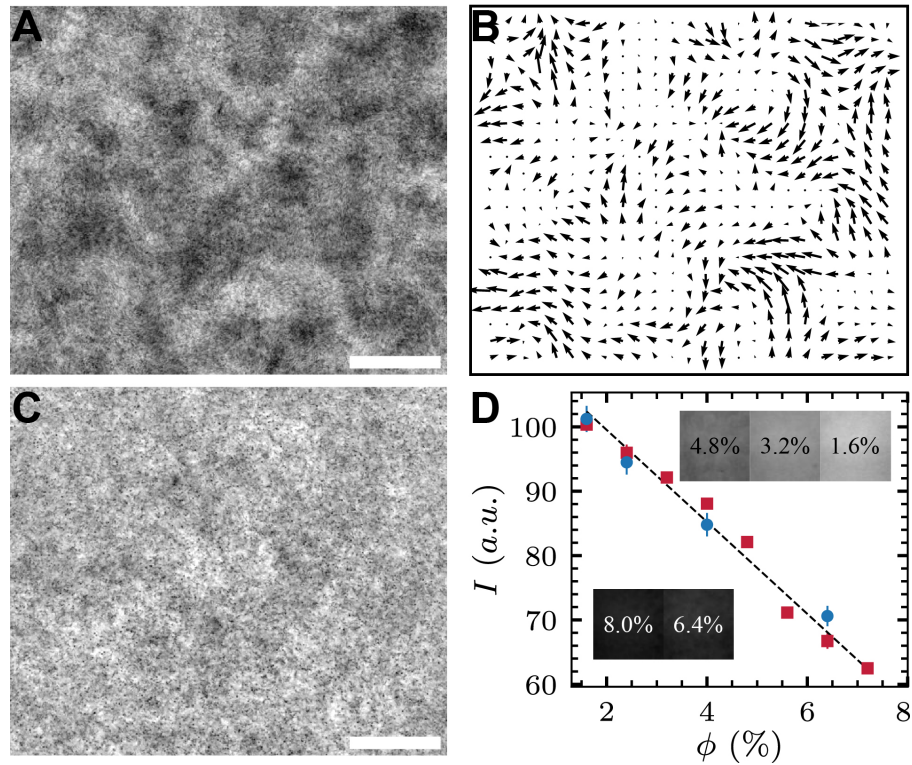

**Fig. 1** Density fluctuations and active turbulence. (A) A snapshot of a dense suspension of *E. coli* with bacterial volume fraction $\phi = 6.4\%$. (B) 2D velocity field of the suspension shown in (A), which exhibits characteristic active turbulent flow patterns. See also Supplementary Video 1. (C) A snapshot of a dilute suspension of *E. coli* with $\phi = 1.6\%$, which shows no active turbulent flows. Scale bars are 85 μm. (D) Average pixel intensities of images, *I*, as a function of $\phi$. Squares indicate the intensity of the spatial average over the full field of view, where disks indicates the temporal average intensity of single pixels. Inset shows the images of bacterial suspensions of different $\phi$ under the same light illumination.

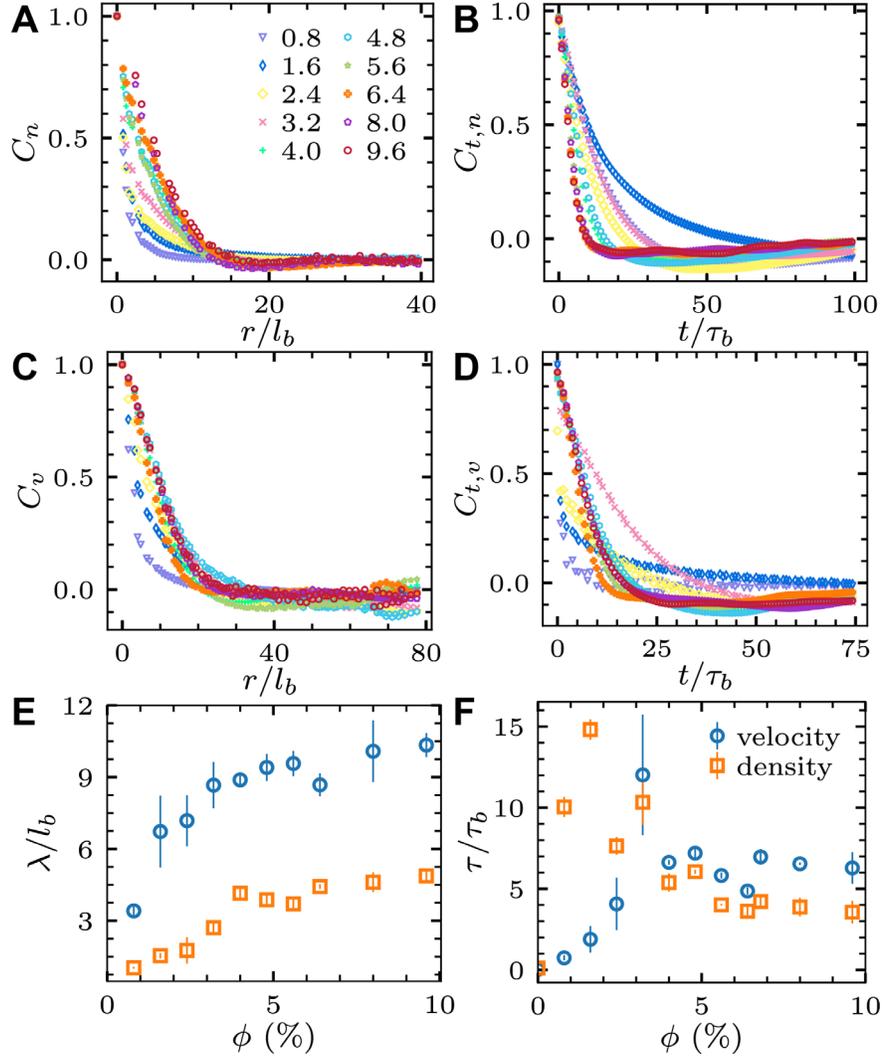

**Fig. 2** Density and velocity spatiotemporal correlations. (A) Density spatial correlations at different bacterial volume fractions $\phi$. Radial position $r$ is normalized by the average length of bacteria $l_b = 3$ μm. (B) Density temporal autocorrelations at different $\phi$. Time $t$ is normalized by the characteristic swimming time of bacteria $\tau_b = 0.2$ s. (C) Velocity spatial correlations at different $\phi$. (D) Velocity temporal autocorrelations at different $\phi$. $\phi$ (%) of different curves are indicated in (A). (E) Density and velocity correlation lengths, $\lambda$, versus $\phi$. (F) Density and velocity correlation times, $\tau$, versus $\phi$. The error bars in (E) and (F) represent the standard deviations of measurements over 3 independent experiments.

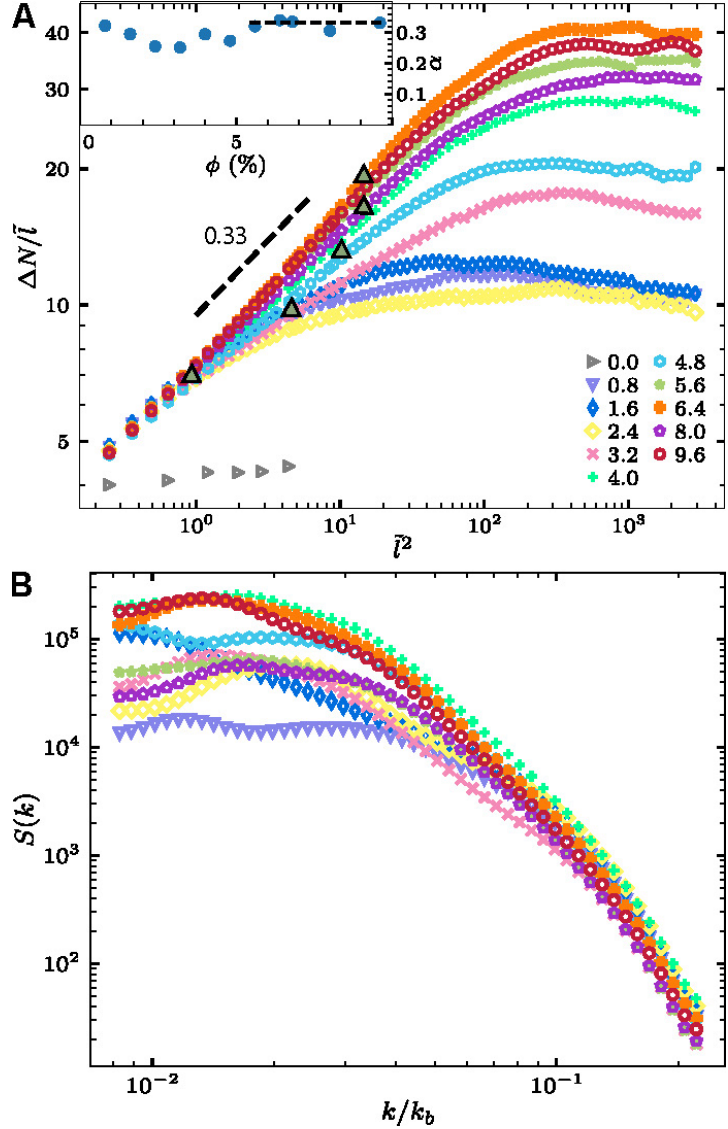

**Fig. 3** Density fluctuations of bacterial suspensions. (A) The standard deviation of bacterial number $\Delta N$ in a subsystem of side length $l$ as a function of the area of the subsystem $l^2$ for suspensions of different volume fractions $\phi$. $\Delta N$ is normalized by the length of the subsystem $l$, which is proportional to the square root of bacterial number in the subsystem $\sqrt{N}$. $l$ is presented in a dimensionless form, $\tilde{l} = l/l_b$, where $l_b = 3$ μm is the average length of bacteria. Dark green triangles indicate the density correlation lengths $\lambda(\phi)$ from Fig. 2E. The black dashed line indicates a power-law scaling of 0.33. The data at $\phi = 0$ measure the temporal fluctuation of the intensity of the microscope illumination light as well as the image noise of the sCMOS camera. Inset: Scaling exponent $\alpha$ versus $\phi$. $\alpha$ is extracted by fitting the experimental data for $l < \lambda(\phi)$. As guide for eye, the dashed line in the inset indicates the constant $\alpha = 1/3$. (B) The Fourier transform of the Ursell function, $S(k)$. The wavenumber $k$ is normalized by $k_b = 2\pi/l_b$, where $l_b = 3$ μm is the length of bacterial body. $S(k)$ is equivalent to the static structure factor for any $k \neq 0$.

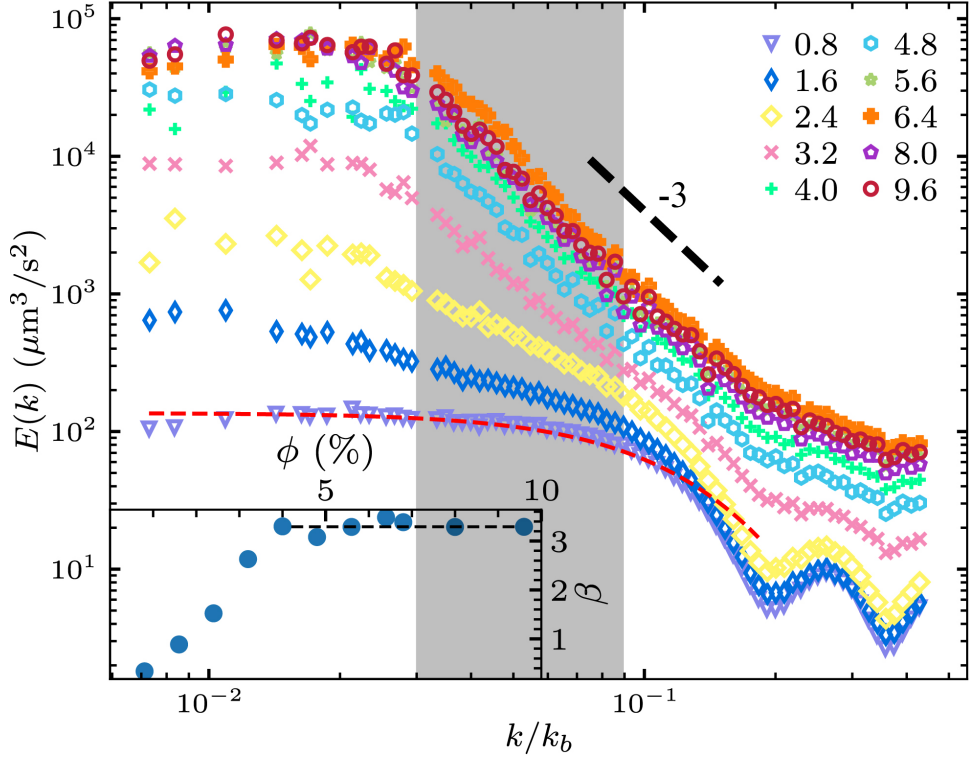

**Fig. 4** Energy spectra $E(k)$ of bacterial suspensions. The wavenumber $k$ is normalized by $k_b = 2\pi/l_b$, where $l_b = 3$ μm is the length of bacterial body. The volume fractions of different suspensions $\phi$ are shown in the legend. Shaded region indicates the range over which the scaling exponent $\beta$ is fitted. The black dashed line shows a power-law scaling of -3. The red dashed line is a fitting of $E(k)$ at $\phi = 0.8\%$ using Eq. (2). In the fitting, the bacterial number density $n$ and the dipole length $l_d = 1.9$ μm are from experiments, whereas the dipole strength $\kappa = 100$ μm$^3$/s and the regularization length $\varepsilon = 14$ μm are taken as fitting parameters. Inset: Scaling exponent of $E(k)$, $\beta$, as a function of $\phi$. The dashed line indicates $\beta = 3.3$.

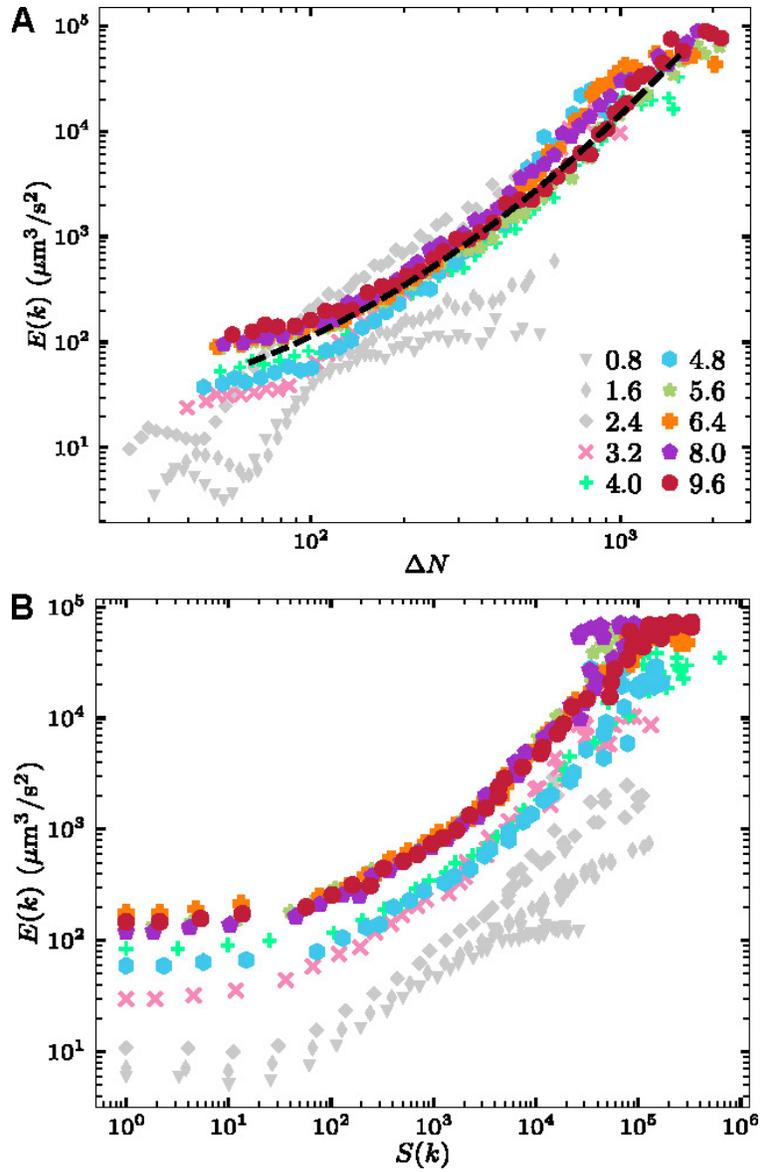

**Fig. 5** Correlation between density fluctuations and energy spectra in the steady state. (A) Energy spectra $E(k)$ plotted against the number density fluctuations $\Delta N$ at the corresponding length scale for bacterial suspensions of different volume fractions $\phi$. Gray symbols are for low-$\phi$ suspensions. The black dashed line is a polynomial fitting of the master curve, serving as guide for the eye. (B) $E(k)$ plotted against the Fourier transform of the Ursell function $S(k)$.

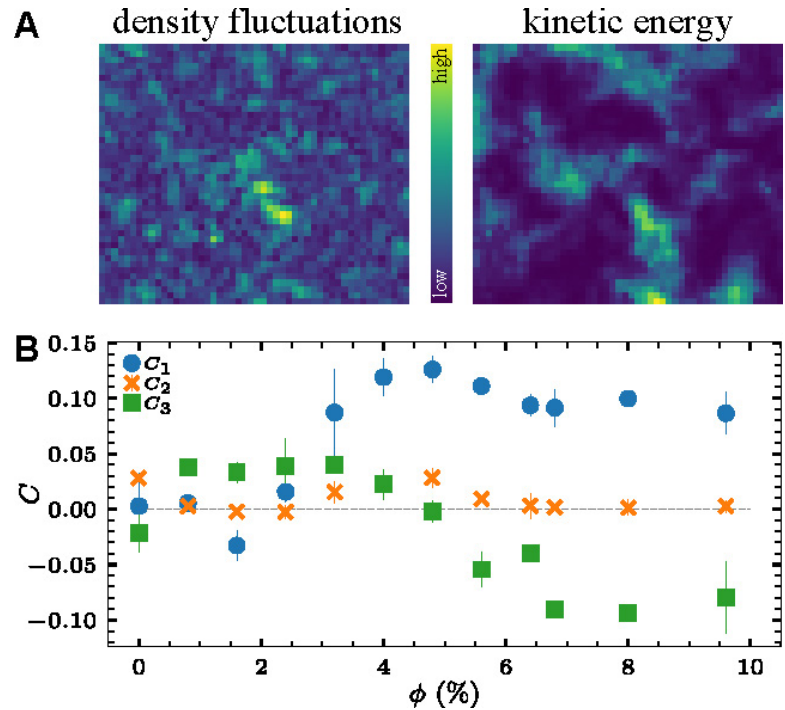

**Fig. 6** Correlation between local density fluctuations and kinetic energies in the steady state. (A) The fields of density fluctuations and kinetic energy at the length scale $l = 2.75 l_b$ in a bacterial suspension of $\phi = 4.8\%$. $l_b = 3$ μm is the average length of bacteria. The field of view is 420 μm by 360 μm. (B) The $\phi$ dependence of the correlation between the local density fluctuations and kinetic energy $C_1$ (disks), the correlation between the local density and the divergence of the bacterial flux $C_2$ (crosses), and the correlation between the local density and kinetic energy $C_3$ (squares). The results are averaged over 1000 frames in the steady state, and the error bars represent the standard deviations.

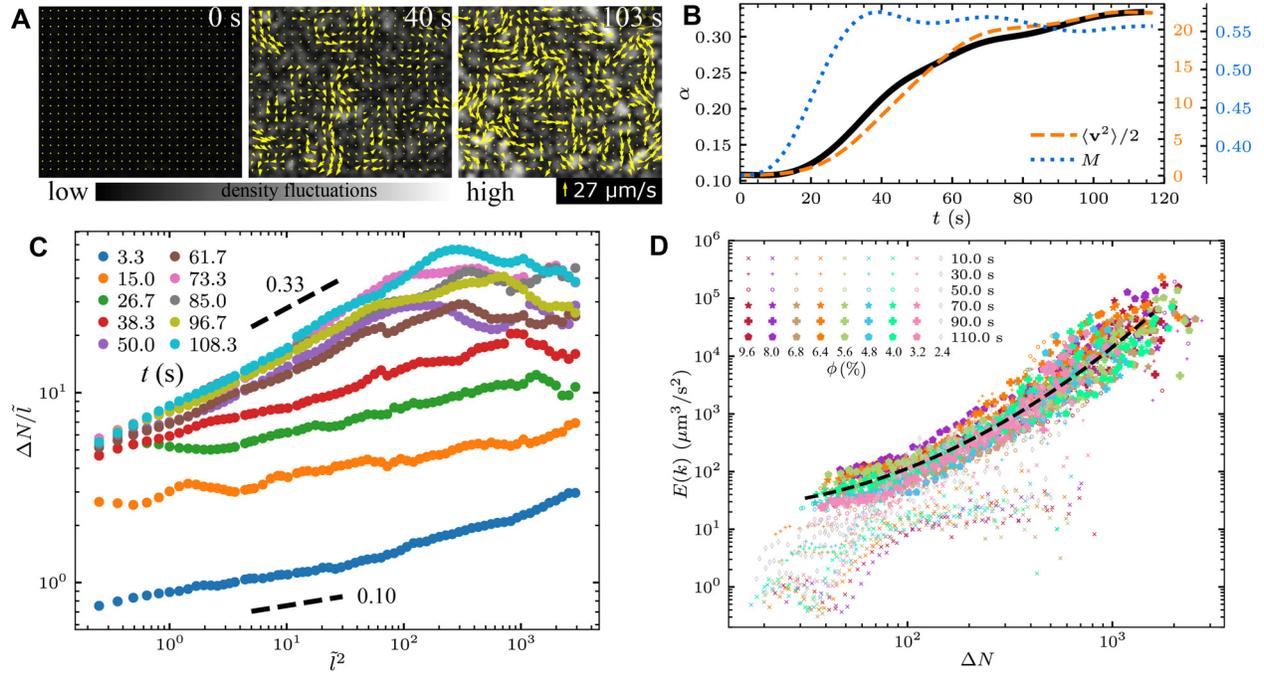

**Fig. 7** Correlation between density fluctuations and energy spectra in the transient state. (A) The field of density fluctuations and the field of the velocity of a bacterial suspension of volume fraction $\phi$ = 6.4% at different time $t$ in the transient state. The gray-scale background indicates the magnitude of local density fluctuations, whereas the yellow arrows show the local velocities. Light is turned on at $t$ = 0. See also Supplementary Video 2. (B) Temporal evolution of the GNF scaling exponent $\alpha$, the total kinetic energy $\langle v^2 \rangle/2$ and the area fraction of regions with collective flow $M$ during the transition to turbulence shown in (A). (C) Number density fluctuations $\Delta N$ as a function of subsystem size $l^2$ at different times over the turbulent transition shown in (A). $\Delta N$ is normalized by the length of subsystems $l$, similar to that in Fig. 3A. $\tilde{l} = l/l_b$. $t$ is indicated in the legend. (D) Energy spectra $E(k)$ versus density fluctuations $\Delta N$ during the turbulent transition. $t$ and $\phi$ are encoded by the marker size and color, respectively. The black dashed line is the master curve shown in Fig. 5A.